\documentclass[12pt]{iopart}

\usepackage{iopams}

\usepackage{graphicx}
\usepackage{subfig}

\usepackage{array}
\usepackage{multirow}

\usepackage{color}

\newcommand{\ket}[1]{|#1\rangle}

\begin{document}

\title{Multi-user quantum key distribution with entangled photons from an AlGaAs chip}

\author{C Autebert$^1$, J Trapateau$^2$, A Orieux$^2$, A Lema\^itre$^3$, C Gomez-Carbonell$^3$, E Diamanti$^2$, I Zaquine$^2$ and S Ducci$^1$}

\address{$^1$ Laboratoire Mat\'eriaux et Ph\'enom\`enes Quantiques, Universit\'e Paris Diderot, Sorbonne Paris Cit\'e, CNRS-UMR 7162, 75205 Paris Cedex 13, France}
\address{$^2$ Laboratoire Traitement et Communication de l'Information, CNRS, T\'el\'ecom ParisTech, Universit\'e Paris-Saclay, 75013, Paris, France}
\address{$^3$ Centre de Nanosciences et de Nanotechnologies, CNRS/Universit\'e Paris Sud, UMR 9001, 91460 Marcoussis, France}
\ead{sara.ducci@univ-paris-diderot.fr}

\vspace{10pt}
\begin{indented}
\item July 2016
\end{indented}

\begin{abstract}
In view of real world applications of quantum information technologies, the combination of miniature quantum resources with existing fibre networks is a crucial issue. Among such resources, on-chip entangled photon sources  play a central role for applications spanning quantum communications, computing and metrology. Here, we use a semiconductor source of entangled photons operating at room temperature in conjunction with standard telecom components to demonstrate multi-user quantum key distribution, a core protocol for securing communications in quantum networks. The source consists of an AlGaAs chip emitting polarization entangled photon pairs over a large bandwidth in the main telecom band around 1550 nm without the use of any off-chip compensation or interferometric scheme; the photon pairs are directly launched into a dense wavelength division multiplexer (DWDM) and secret keys are distributed between several pairs of users communicating through different channels. We achieve a visibility measured after the DWDM of 87\% and show long-distance key distribution using a 50-km standard telecom fibre link between two network users. These results illustrate a promising route to practical, resource-efficient implementations adapted to quantum network infrastructures.
\end{abstract}

%%%%%%%%%%%%%%%%%%%%%%%%%%%%%%%%%%%%%%%%%%%%%%%%%%%%%%%%%%%%%%%%%%%%%%

\section{Introduction}

Entangled photon pairs are an important resource in quantum information including quantum communications and computing \cite{Horodecki2009,Scarani2009,Brunner2014}. In particular for quantum communications, photons are the most suitable information carriers because of their high transmission speed and their robustness to decoherence. To enable the wide use of entangled photon pairs in future quantum telecommunication networks, practical sources of photonic entanglement combined with techniques allowing to reduce the required quantum resources are highly desirable. The motivation is therefore strong to develop compact, high-performance sources operating at room temperature that can be easily fabricated and integrated into telecom fiber networks. In this context, semiconductor materials play a central role; indeed, great progress has been reported in last years on quantum devices based, for instance, on the Silicon and AlGaAs integration platforms~\cite{JOpt2016Review}. The latter is particularly appealing for effective miniaturization, thanks to the direct bandgap and strong electro-optical effect exhibited by AlGaAs. Moreover, the small birefringence of AlGaAs waveguides has led to the demonstration of polarization entangled photon states without any off-chip compensation~\cite{Orieux2013,Horn2013,Helmy1511}, which provides a significant advantage towards achieving stable and scalable architectures with a minimum amount of components.

At the same time, techniques widely used in classical telecommunications \cite{Winzer2015}, such as dense wavelength division multiplexing (DWDM) and active optical routing, have been shown to be compatible with entangled photon distribution \cite{Lim2008,Poppe2013,Ghalbouni2013,Trapateau2015,Qian1506,Tanzilli2016}, and hence to effectively reduce the number of required sources in a network. Indeed, with the use of DWDM it is possible to distribute entangled photon pairs emitted in a large band to several user pairs at the same time, while using only one source~\cite{Trapateau2015,Ciurana2015}. Such schemes can then be made fully reconfigurable by adding active optical switches allowing to distribute an entangled photon pair between any pair of users in the network; this technique has been used recently with PPLN waveguides~\cite{Poppe2013,Tanzilli2016} and with a periodically poled fiber~\cite{Qian1506}.

In this work, we bring together such elements to demonstrate in practice one of the most important quantum communication protocols, namely quantum key distribution (QKD), in a network setting. We show in particular the distribution of quantum secret keys between multiple pairs of users using an AlGaAs source emitting broadband entangled photons and standard telecom components for wavelength division multiplexing. It is important to remark that although QKD can be implemented with single photons or coherent light pulses \cite{Scarani2009}, entanglement-based QKD \cite{Ekert91,BBM92} is necessary to achieve device independence \cite{Acin2007}, hence removing security loopholes due to imperfections of practical components. In our implementation, the key information is encoded on polarization entangled photons~\cite{Qian1506,NJPsecocq}, which are readily produced by our source. This allows us to use a simple experimental setup that does not necessitate stabilized interferometers required for systems based on time-bin or energy-time entanglement~\cite{Tanzilli2016}. With this setup we obtain high visibility and low quantum bit error rate results for several pairs of users separated by up to 50 km of standard optical fiber.

\section{Experimental setup and methods}

\begin{figure}[!htbp]
	\centering
	\includegraphics[width =0.9\linewidth]{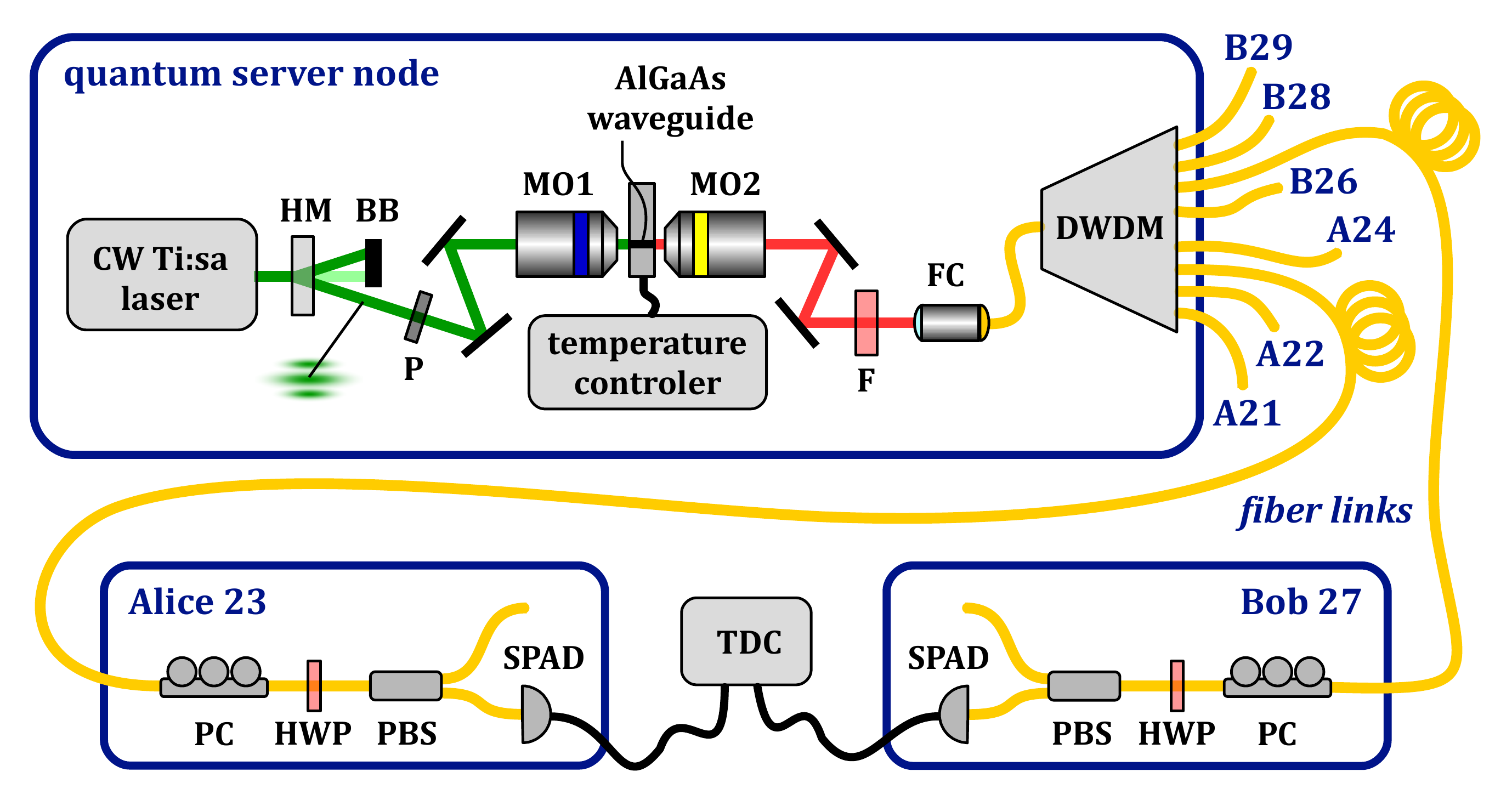}
	\caption{Sketch of the multi-user entangled-based quantum key distribution setup. HM: holographic mask; BB: beam block; P: polarizer; MO1, MO2: microscope objectives; F: longpass spectral filter; FC: fiber coupler; DWDM: dense wavelength demultiplexer; PC: polarization controller;  HWP: half-wave plate; PBS: fibered polarizion beam splitter; SPAD: single-photon avalanche photodiode; TDC: time-to-digital converter.}
	\label{fig1}
\end{figure}

The schematic of our experimental setup is shown in Fig.~\ref{fig1}. This setup implements the BBM92 QKD protocol~\cite{BBM92}, which is the entanglement based generalization of the standard BB84 protocol \cite{BB84}. In this protocol, two distant users, Alice and Bob, share entangled photon pairs in a Bell state generated by a source that can be situated in either Alice's or Bob's site or in the middle, for instance at a server node in a telecommunication network. This node is the central element of our setup and includes the entangled photon source and a demultiplexing device, as shown in Fig.~\ref{fig1}.

\begin{figure}[!htbp]
	\centering
	\subfloat[]{\includegraphics [height =0.35\linewidth]{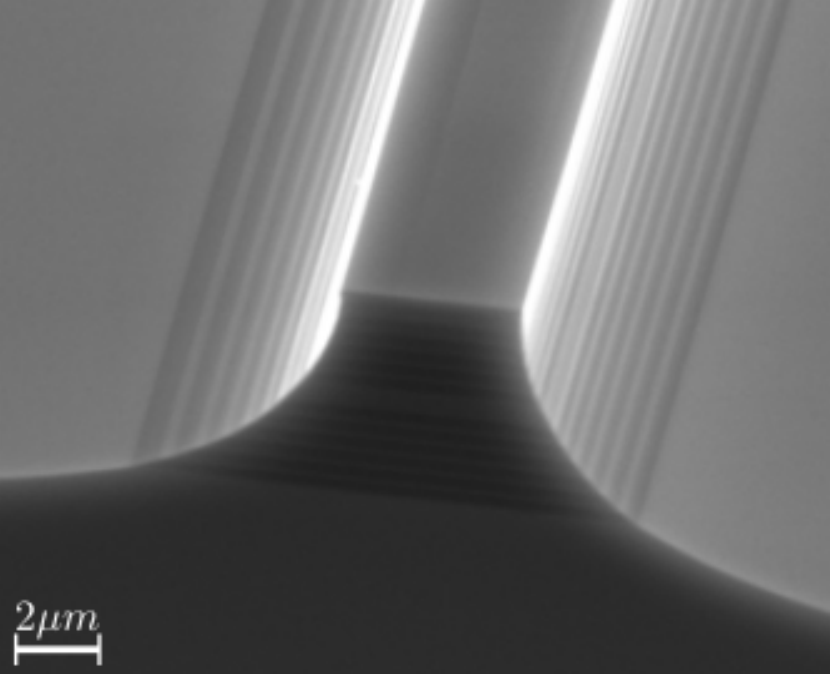}} \hspace{0.1cm}
	\subfloat[]{\includegraphics [height =0.35\linewidth]{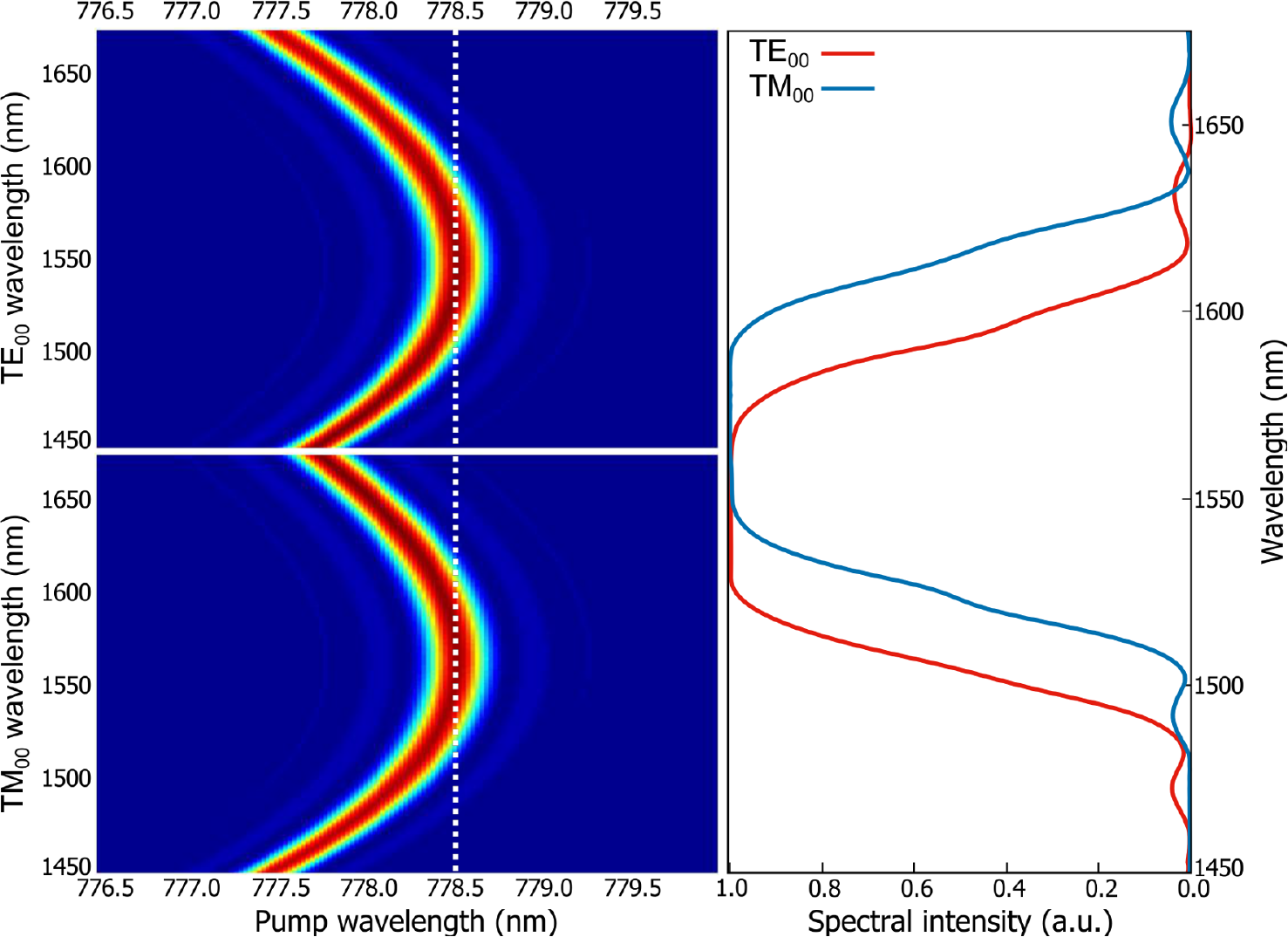}}
	\caption{(a) Scanning Electron Microscope image of the sample: the upper and lower Bragg mirrors as well as the waveguide core are clearly visible. (b) Simulated tuning curves of the type-II phase matching at $T=20^\circ$C.}
	\label{fig2}
\end{figure}

The source consists of a multilayer AlGaAs ridge waveguide (see Fig.~\ref{fig2}(a)), designed to emit photon pairs through type-II spontaneous parametric down conversion (SPDC)
(see Supplementary Information for more details on the sample). The modes involved in the nonlinear process are a TE Bragg mode at 779~nm for the pump~\cite{Yeh1976,Helmy2006,Autebert2016} and TE$_{00}$ and TM$_{00}$ modes around the degeneracy wavelength of
1.56~$\mu$m for the photon pair. The TE Bragg mode of the sample is excited by using the light beam of a continuous-wave (CW) Ti:sapphire laser %(SolsTiS M Squared)
that is first spatially shaped with a holographic mask (HM) and then focused to the sample through a microscope objective (MO1). The wavelengths of the emitted photon pairs are determined by the following energy conservation and phase-matching relations:
\begin{eqnarray}
	\hbar\omega_{p} = \hbar\omega_{A}+\hbar \omega_{B} \\
	n_{\mathtt{\,TE}_{Bragg}}(\omega_{p})\,\omega_{p}
	= n_{\mathtt{\,TE}_{00}}(\omega_{A})\, \omega_{A} + n_{\mathtt{\,TM}_{00}}(\omega_{B})\,\omega_{B}\\
	n_{\mathtt{\,TE}_{Bragg}}(\omega_{p})\,\omega_{p}
	= n_{\mathtt{\,TM}_{00}}(\omega_{A})\, \omega_{A} + n_{\mathtt{\,TE}_{00}}(\omega_{B})\,\omega_{B},
\label{equation1}
\end{eqnarray}

\noindent where $\omega_{A}$ and $\omega_{B}$ (with $\omega_{A}\leq\omega_{B}$) are the optical angular frequencies of the down-converted photons, $\omega_{p}$ is the pump frequency, and $n_{i}$ (with $i=\mathrm{TE}_{Bragg},$ $\mathrm{TE}_{00},\mathrm{TM}_{00}$) are the effective indices of the three guided modes.
The simulated tuning curves for a 2.1~mm-long sample pumped by a CW laser at a temperature of 20$^\circ$C are shown in Fig.~\ref{fig2}(b); in these simulations we have modeled the refractive index of the materials using Refs~\cite{chilwell_thin-films_1984,gehrsitz_refractive_2000}. For a given pump wavelength, the pairs emitted over the entire region of spectral overlap between the $\mathrm{TE}_{00}$ and $\mathrm{TM}_{00}$ modes are produced in the polarization entangled Bell state
\begin{equation}
	\ket{\Psi} = \frac{1}{\sqrt{2}}(\ket{01} + \ket{10})_{A,B}
\label{equationBell}
\end{equation}
where the logical states $\ket{0}$ and $\ket{1}$ correspond to the polarization modes $\mathrm{TE}_{00}$ and $\mathrm{TM}_{00}$, respectively, and $A$ ($B$) corresponds to the longer (shorter) wavelength photon of the pair. Note that due to the very small birefringence ($\Delta n\approx 7\times10^{-3}$) of the AlGaAs waveguide, the group velocity mismatch between the photons of a pair is so small that no off-chip compensation of the walk-off is required to obtain the state of Eq.~(\ref{equationBell})~\cite{Orieux2013,Horn2013}. This is a key aspect that enables the direct use of the generated pairs and therefore the easy integration of our device in simple and robust architectures for practical applications.

As shown in Fig.~\ref{fig1}, light emerging from the opposite end of the sample is collected with a second microscope objective (MO2) and sent to a fibre coupler (FC), after filtering out the pump wavelength with a longpass spectral filter (F). The entangled photon pairs are then sent to four different pairs of users through the output channels of an 8-channel dense wavelength demultiplexer (DWDM) belonging to the 100~GHz International Telecommunication Union (ITU) grid (100~GHz channel width and 100~GHz channel separation). We remark that when a narrowband pump laser is used, the strong anti-correlation between the frequencies of the photons of a given pair makes them exit through symmetric frequency channels with respect to the degeneracy frequency $\omega_{p}/2$. Hence, the working point of our device is set such that the phase-matching resonance corresponds to the central channel of the DWDM (OptoLink) %OLDWDM-SD-1-16-22-1-FU-D model)
used in our QKD setup; in particular, the temperature of the sample is adjusted with a Peltier cooler while monitoring the coincidence to accidental (CAR) ratio and the emission wavelength, which allows us to find an optimum working point at a temperature $T= 19.5^\circ$C. The targeted central channel is channel 25 of the ITU grid, while channels 21-24 and 26-29 are sent to Alice and Bob, respectively. The choice of the OptoLink dielectric thin film DWDM was the result of an analysis of the performance of various demultiplexer technologies with respect in particular to their effect on the distribution of polarization entangled photon pairs~\cite{Trapateau2015}.

The photons of a given pair of users then exit the central node of our setup and either directly enter Alice's and Bob's sites or first propagate through a 25-km spool of standard single mode fibre each, leading to a 50-km total fiber link between the users. The stations of Alice and Bob contain a polarization controller (PC), which is required because our DWDM is a non polarization-maintaining device, followed by a polarization analysis setup including a half-wave plate (HWP), a fibred polarization beam-splitter (PBS) and an InGaAs free-running single-photon detector (IdQuantique ID220). Finally, the detectors of Alice and Bob are connected to a time-to-digital converter (TDC) in order to perform coincidence measurements.

These setups for Alice and Bob partly realize the required steps for the BBM92 protocol, where upon reception of the entangled photon pairs, the two communicating parties both locally measure their respective photon, choosing randomly and independently among the computational basis $\{\ket{0},\ket{1}\}$ or the diagonal basis $\{\ket{+},\ket{-}\}$ (with $\ket{\pm}=(\ket{0}\pm\ket{1})/\sqrt{2}$). After a basis reconciliation step and post-processing procedures, including error correction and privacy amplification, performed through a classical communication channel, they can extract a secret key from their coincidence measurements in identical bases. We note here that for practical reasons in our experiment we measured coincidences separately in the eight measurement configurations where identical bases are chosen by Alice and Bob. The full implementation of the random basis choice could be achieved passively by means of a 50-50 beam-splitter routing the photon to a polarization beam-splitter aligned either with the computational or the diagonal basis, and four single-photon detectors for each user. Additionally, only a single detector per user is required if temporal multiplexing of the four projected polarization states can be implemented after the polarization beam-splitters~\cite{Qian1506}.

At a high level, the information-theoretic security of the key distribution process is guaranteed as long as the entanglement shared by Alice and Bob is sufficiently high. More specifically, from the visibility of their coincidence measurements, Alice and Bob can estimate the quantum bit error rate (QBER), $e$, of their quantum communication link~\cite{NJPTenerife}:
\begin{equation}
	e = \frac{1-V_{\mathrm{tot}}}{2},
\label{equationQBER}
\end{equation}
where $V_{\mathrm{tot}}$ is the total entanglement visibility in both the computational and the diagonal bases:
\begin{equation}
	V_{\mathrm{tot}} = \frac{C_{\mathrm{max}}-C_{\mathrm{min}}}{C_{\mathrm{max}}+C_{\mathrm{min}}},
\label{equationV}
\end{equation}
where $C_{\mathrm{max}}$ and $C_{\mathrm{min}}$ are respectively the maximum and minimum number of coincidences observed in both bases.

From this value of QBER and the total rate of coincidence counts, $R_{\mathrm{raw}}$, measured in all possible combinations of both bases, Alice and Bob can then estimate a lower bound for their secret key rate, $R_{\mathrm{key}}$~\cite{MaFungLo2007}:
\begin{equation}
	R_{\mathrm{key}} \geq \frac{1}{2}R_{\mathrm{raw}}\left( 1 - f(e)H_2(e) - H_2(e) \right),
\label{equationRkey}
\end{equation}
where the $1/2$ prefactor is the basis reconciliation (or sifting) factor that accounts for the fact that only the measurements where Alice and Bob choose the same basis can be used for the key, which happens half of the time; the quantity $f(e)$ corresponds to the efficiency of the error-correction code, where in general $f(e) \geq 1$ and the Shannon limit is $f(e) = 1$; and $H_2$ is the binary entropy function, $H_2(x) = - x\log_2(x) - (1-x)\log_2(1-x)$. The bound of Eq.~(\ref{equationRkey}) is derived against general coherent attacks~\cite{MaFungLo2007}, hence providing maximal security guarantees.

\section{Results}

Using the experimental setup and techniques described above, we first performed QKD experiments with the entangled photons entering directly the stations of Alice and Bob, for all four symmetric channel pairs, namely 21-29; 22-28; 23-27; 24-26, of the DWDM, corresponding to four different pairs of users sharing a secret key.

\begin{figure}[!htbp]
	\centering
	\centering
	\subfloat[]{\includegraphics [width =0.5\linewidth]{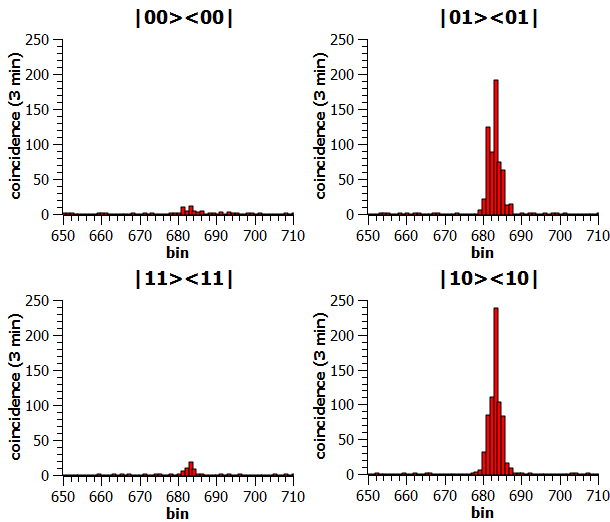}}
	\subfloat[]{\includegraphics [width =0.5\linewidth]{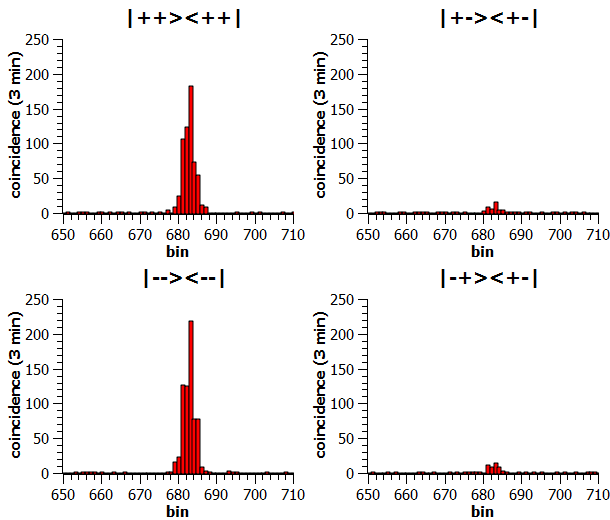}}
	\caption{Coincidence histograms of the eight different projective measurements, corresponding to an identical choice of basis by Alice and Bob, used to estimate the QBER and the secret key generation rate for channels 23 and 27 centered at 1558.98~nm and 1555.75~nm, respectively, for zero distance between Alice and Bob. Data was accumulated during 180~s with a sampling resolution of 164~ps. (a) both Alice and Bob choose the computational basis; (b) they both choose the diagonal basis.}
	\label{fig3}
\end{figure}

In Fig.~\ref{fig3} we show the measured coincidence histograms corresponding to the eight possible projective measurements obtained when Alice and Bob make the same basis choice, for the 23-27 channel pair. From these measurements, we can estimate in all cases the sifted key generation rate, $R_{\mathrm{sift}} = \frac{1}{2}R_{\mathrm{raw}}$, and the total entanglement visibility, $V_{\mathrm{tot}}$ (see Supplementary Information for details). The latter leads to the calculation of the QBER (Eq.~(\ref{equationQBER})) and finally taking standard values for $f(e)$~\cite{NJPTenerife,Yamamoto2002}, we can use Eq.~(\ref{equationRkey}) to estimate the secret key rate, $R_{\mathrm{key}}$. For the 23-27 channel pair, we obtain $R_{\mathrm{sift}} = 13.8 \pm 0.3$~bit/s, $V_{\mathrm{tot}} = 86.7\% \pm 1.0\%$, $e = 0.066 \pm 0.005$, and $R_{\mathrm{key}} = 3.28 \pm 0.08$~bit/s. The results for all different pairs of users are summarized in Table~\ref{table1}.

\begin{table*}[!htbp]
\begin{center}
\begin{tabular}{ |c | c || c | c || c | c | }
\hline
channel & fibre link & $V_{\mathrm{tot}}$ & $R_{\mathrm{sift}}$ & QBER & $R_{\mathrm{key}}$ \\
pair & distance & (\%) & (bit/s) & (\%) & (bit/s)\\
\hline
\hline
21--29 & 0 km & 90.0 $\pm$ 1.0 & 9.91 $\pm$ 0.23 & 5.0 $\pm$ 0.5 & 3.79 $\pm$ 0.10 \\ \hline
22--28 & 0 km & 83.5 $\pm$ 2.1 & 13.0 $\pm$ 0.27 & 8.2 $\pm$ 1.0 & 1.31 $\pm$ 0.09 \\ \hline
\multirow{2}{*}{23--27} & 0 km & 86.7 $\pm$ 1.0 & 13.8 $\pm$ 0.3 & 6.6 $\pm$ 0.5 & 3.28 $\pm$ 0.08 \\
    & 50 km & 86.1 $\pm$ 2.9 & 1.01 $\pm$ 0.06 & 6.9 $\pm$ 1.5 & 0.21 $\pm$ 0.13 \\ \hline
24--26 & 0 km & 87.9 $\pm$ 1.4 & 6.9 $\pm$ 0.2 & 6.1 $\pm$ 0.7 & 1.95 $\pm$ 0.08 \\ \hline
\end{tabular}
\caption{Measured and estimated quantities for the QKD protocol for four different DWDM channel pairs centered on channel 25. Errors are calculated assuming Poisson statistics for the coincidence counts. Finite-size effects have not been considered in these results.}
\label{table1}
\end{center}
\end{table*}

In order to assess the performance of our system over long distances, we repeat the experiments, for the 23-27 channel pair, with a spool of 25-km standard single-mode fibre inserted after the DWDM in both user paths, leading to a total distance of 50~km between Alice and Bob. The measured coincidence histograms in this case are shown in the Supplementary Information. In this configuration, the QBER increases slightly to $e = 0.069 \pm 0.015$, and the secret key rate we obtain is $R_{\mathrm{key}} = 0.21 \pm 0.13$~bit/s. These results are also included in Table~\ref{table1}.

\begin{figure}[!htbp]
	\centering
	\includegraphics[width = 0.6\linewidth]{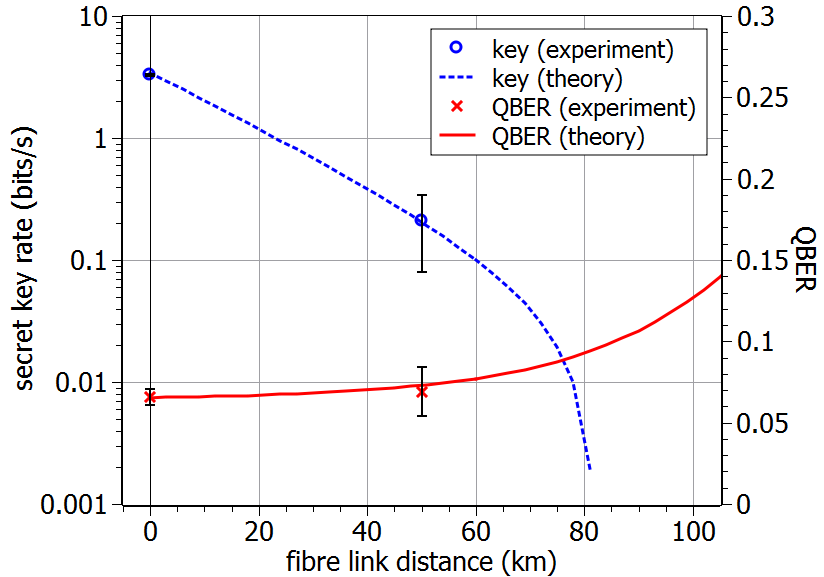}
	\caption{Evolution of the secret key generation rate and the QBER as a function of the distance between Alice and Bob for the 23-27 channel pair. The theoretical fits are based on the model of Ref \cite{Yamamoto2002} and take into account the quantum efficiency and dark counts of the single-photon detectors, the collection efficiency of the setup, standard loss of single-mode fibers and a baseline polarization measurement error; the values of these experimental parameters are provided in the Supplementary Information.}
	\label{figVSdistance}
\end{figure}

The ultimate long-distance performance of our QKD system is illustrated in Fig.~\ref{figVSdistance}, where we show the two experimental points of the secret key generation rate and the QBER as a function of the distance for the 23-27 channel pair, as well as a theoretical fit derived from a model of our experiment taking into account all experimental parameters; this model is based on Ref \cite{Yamamoto2002} and is detailed in the Supplementary Information. We see that secret key generation is possible for a maximal distance of around 80~km in our experimental conditions.

\section{Discussion}

In Table~\ref{table2}, we provide representative results obtained in entanglement-based QKD experiments over standard single-mode fiber links in the telecom C band together with some of our results.
A direct comparison between the various experiments in not straightforward as the experimental conditions may vary or may be described with respect to different parameters, such as optimal pump power \cite{NJPsecocq}, mean photon number per detection window per channel pair \cite{Tanzilli2016}, etc. It is clear, however, that these first multi-user entanglement-based QKD results, obtained with a semiconductor chip source, are very promising for quantum communication protocol implementations based on a particularly simple experimental configuration.

\begin{table*}[htbp]
\begin{center}
\begin{tabular}{ |c || c | c || c || c | c || c | c | }
\hline
Ref & \multicolumn{2}{c||}{\cite{NJPsecocq}} & \cite{Qian1506} & \cite{Tanzilli2016} & \multicolumn{2}{c|}{this work} \\
\hline
\hline
source & \multicolumn{2}{c||}{PPKTP crystal} & poled fiber & PPLN WG & \multicolumn{2}{c|}{AlGaAs WG} \\
 & \multicolumn{2}{c||}{type~I} & type~II & type~0 & \multicolumn{2}{c|}{type~II} \\ \hline
qubit type & \multicolumn{2}{c||}{polarization} & polarization & energy-time & \multicolumn{2}{c|}{polarization} \\ \hline
pumping regime & \multicolumn{2}{c||}{CW} & 81.6~MHz & CW & \multicolumn{2}{c|}{CW} \\ \hline
SPAD & \multicolumn{2}{c||}{ID201$^\dagger$} & ID220 & ID220 \& ID230 & \multicolumn{2}{c|}{ID220} \\ \hline
DWDM grid & \multicolumn{2}{c||}{3~nm$^{\dagger\dagger}$} & 200~GHz & 100~GHz & \multicolumn{2}{c|}{100~GHz} \\ \hline
\hline
distance (km) & 0 & 16$^{\dagger\dagger\dagger}$ & 40 & 150 & 0 & 50 \\ \hline
QBER & 2.3\% & 3.5\% & 2.0\% & 6.0\%$^\ddagger$ & 6.6\% & 6.9\% \\ \hline
$R_{\mathrm{key}}$ (bits/s) & 12$\times10^3$ & 2$\times10^3$ & 20 & 0.3$^\ddagger$ & 3.3 & 0.21 \\ \hline
\end{tabular}
\caption{Relevant state-of-the art entanglement-based QKD experiments over standard single-mode fiber links in the telecom C band.\\
$\dagger$: Bob's photon only (Alice's photon at 810~nm was detected close to the source by a silicium SPAD). $\dagger\dagger$: no DWDM was used, the photons were filtered with a 3~nm bandpass filter. $\dagger\dagger\dagger$: installed fiber link in Vienna. $\ddagger$: estimated from reported coincidence rates and visibilities.}
\label{table2}
\end{center}
\end{table*}

Further progress is possible in multiple ways. The design of the source can be optimized and the global impact of this improvement on our QKD system can be readily evaluated:
the collection efficiency is presently limited by the semiconductor facet reflectivity (25\%) and the mode mismatch between the semiconductor waveguide and the single-mode fiber (25\% coupling efficiency); the use of an antireflection coating and of a laser-diode-to-fiber-coupler %(e.g. OZoptics DTS0063)
for the coupling of the emitted photons to the fiber could improve the collection efficiency by a factor 4.2, reaching 21\% instead of the current value of 5\%.
Moreover, the use of commercially available superconducting detectors with a quantum efficiency of 87\% and a dark count rate of 10 count/s (see for instance Ref \cite{quantumopus}) would increase the detection efficiency by a factor 4.3, from 20\% to 87\%, and remove the influence of dark counts, leaving only spurious light related false coincidences.
Finally, the polarization-maintaining fibers of our polarization analysis setups introduced polarization-mode dispersion (PMD), which limited the maximum possible visibility to 92\% (see Supplementary Information). Replacing these fibers by standard single-mode fibers could reduce the polarization error rate from the current value of 6\% to 2.5\%.
Taking into account all these elements, the performance of the setup proposed in this work could reach a secret key generation rate of 2.8~kbit/s in the same pumping conditions at zero distance and could allow key distribution at a distance of up to 230~km between users.

As far as the semiconductor chip is concerned, the fabrication of highly efficient low loss waveguides is currently under intensive investigation; in particular the development of inductively coupled plasma (ICP) processes to etch straight flank waveguides while keeping low optical losses would increase the confinement of the interacting optical fields and consequently the efficiency of the parametric down conversion. Encouraging results have recently been obtained on our devices leading to a CAR on the order of 1700 \cite{Autebert2016inprep}.

In conclusion, we have implemented, for the first time, a setup allowing multi-user QKD using a semiconductor source of entangled photons and standard telecom components. We have performed entanglement-based quantum key distribution between users at a distance of 50~km with a secret key rate of 0.21~bit/s. The compliance of our source with electrical pumping~\cite{Boitier2014}, the possible improvement in its design together with the continuous progress in detector fabrication techniques make our approach a promising candidate for real-world quantum communications.

%%%%%%%%%%%%%%%%%%%%%%%%%%%%%%%%%%%%%%%%%%%%%%%%%%%%%%%%%%%%%%%%%%%%%%
\section*{Acknowledgments}
C. Autebert acknowledges the D\'el\'egation G\'en\'erale de l'Armement (DGA) for funding. We also acknowledge financial support by the Agence Nationale de la Recherche (ANR-14-CE26-0029-01, and projects COMB and QRYPTOS), the R\'egion Ile-de-France in the framework of DIM Nano-K (project QUIN), the French network RENATECH, the City of Paris (project CiQWii), and the France-USA Partner University Fund (project CRYSP). S. Ducci is a member of the Institut Universitaire de France.

%%%%%%%%%%%%%%%%%%%%%%%%%%%%%%%%%%%%%%%%%%%%%%%%%%%%%%%%%%%%%%%%%%%%%%
\section*{Supplementary Information}

\textbf{Source details}. The sample consists of a 6-period Al$_{0.80}$Ga$_{0.20}$As/Al$_{0.25}$Ga$_{0.75}$As Bragg reflector (lower cladding), a 298\,nm Al$_{0.45}$Ga$_{0.55}$As core and a 6-period Al$_{0.25}$Ga$_{0.75}$As/Al$_{0.80}$Ga$_{0.20}$As Bragg reflector (upper cladding). It is grown by molecular beam epitaxy on a (100) GaAs substrate and waveguides are fabricated using wet chemical etching to define 5.5-6~$\mu$m wide and 5~$\mu$m deep ridges along the (011) crystalline axis, in order to use the maximum non-zero optical nonlinear coefficient and a natural cleavage plane.\\

\begin{figure}[!htbp]
	\centering
	\centering
	\subfloat[]{\includegraphics [width =0.5\linewidth]{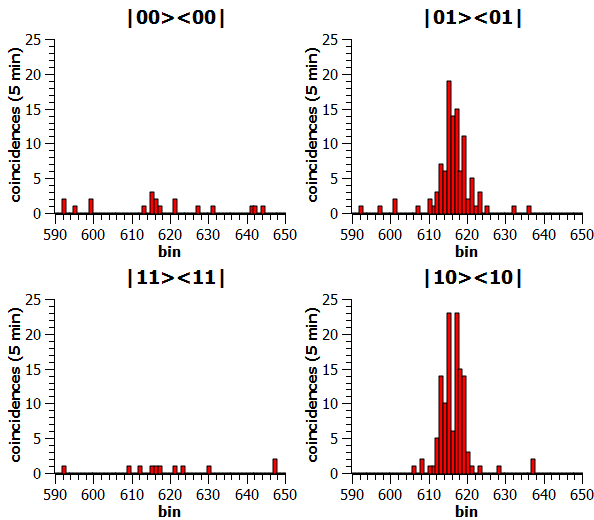}}
	\subfloat[]{\includegraphics [width =0.5\linewidth]{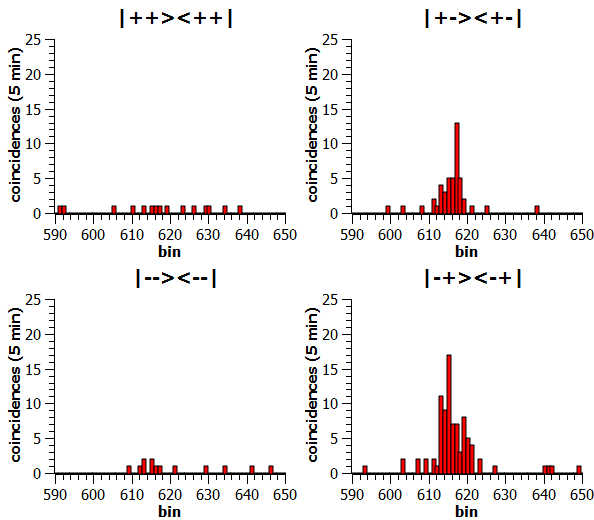}}
	\caption{Coincidence histograms of the eight different projective measurements used to estimate the QBER and the secret key generation rate for the 23-27 channel pair, for a 50~km distance between Alice and Bob. Data was accumulated during 300~s with a sampling resolution of 164~ps. (a) both Alice and Bob choose the computational basis; (b) they both choose the diagonal basis.}
	\label{fig3_50km}
\end{figure}

\noindent \textbf{Measured coincidence data and analysis}. The measured coincidence histograms for the 50-km experiment using the 23-27 DWDM channel pair are shown in Fig.~\ref{fig3_50km}.

\noindent The sifted key generation rate is calculated from the obtained data as follows:
\begin{equation}
	R_{\mathrm{sift}} = (C^{00}_p + C^{01}_p + C^{11}_p + C^{10}_p + C^{++}_p + C^{+-}_p + C^{--}_p + C^{-+}_p)/\tau,
\label{equationRsift}
\end{equation}
where each of the eight terms is obtained from adding together the number of coincidences measured in the five (seven) bins corresponding to the coincidence peak of Fig.~\ref{fig3} (Fig.~\ref{fig3_50km}) accumulated over a time $\tau = 3$~min ($5$~min).\\

\noindent It is also possible to calculate the false coincidence rate as follows:
\begin{equation}
	R_{\mathrm{false}} = 2(C^{00}_0 + C^{01}_0 + C^{11}_0 + C^{10}_0 + C^{++}_0 + C^{+-}_0 + C^{--}_0 + C^{-+}_0)/\tau,
\label{equationRfalse}
\end{equation}
where each of the terms is obtained from the mean value of the number of coincidences measured in five (seven) bins outside the coincidence peak of Fig.~\ref{fig3} (Fig.~\ref{fig3_50km}). As explained below, these values were used to estimate the SPDC generation probability of our setup.\\

\noindent Finally, the maximum and minimum number of coincidences in both bases are given by:
\begin{equation}
	C_{\mathrm{max}} = C^{01}_p + C^{10}_p + C^{++}_p + C^{--}_p
\label{equationCmax}
\end{equation}
\begin{equation}
	C_{\mathrm{min}} = C^{00}_p + C^{11}_p + C^{+-}_p + C^{-+}_p,
\label{equationCmin}
\end{equation}
which are inserted in Eq.~(\ref{equationV}) to calculate $V_{\mathrm{tot}}$. (Note that Eqs.~(\ref{equationCmax}) and (\ref{equationCmin}) only apply if the state seen by Alice and Bob is that of Eq.~(\ref{equationBell}). If the state has a $\pi$ phase shift instead: $\ket{\Psi} = \frac{1}{\sqrt{2}}(\ket{01} - \ket{10})_{A,B}$, as was the case for the experiment with the two spools of 25~km, the terms $C^{++}_p + C^{--}_p$ and $C^{+-}_p + C^{-+}_p$ need to be interchanged.)\\

\noindent \textbf{Theoretical model for secret key generation as a function of distance}. The fitting curves in Fig.~\ref{figVSdistance} were calculated following the model of Ref~\cite{Yamamoto2002}, with the security bound of Ref~\cite{MaFungLo2007} (Eq.~(\ref{equationRkey})). In particular, the error rate is given by the expression:
\begin{equation}
	e = \frac{\frac{1}{2}p_{\mathrm{false}}+b\,p_{\mathrm{true}}}{p_{\mathrm{true}}+p_{\mathrm{false}}},
\label{equationFitQBER}
\end{equation}
where $p_{\mathrm{true}}$ and $p_{\mathrm{false}}$ are the probabilities of true coincidences (originating from photons of the same pair) and accidental coincidences (from noise and lost photons) respectively, and $b$ is a parameter accounting for a systematic polarization measurement error.
The true and false coincidence probabilities in the case of a SPDC source in the CW pumping regime can be expressed as~\cite{Takesue2010}:
\begin{eqnarray}
	p_{\mathrm{true}} &=& \mu \,\eta^2\\
	p_{\mathrm{false}} &=& \mu^2 \eta^2 + 8 \, \mu \, \eta \, d + 16 \, d^2,
\label{equationPtPf}
\end{eqnarray}
where $\mu$ is the mean number of photons per pulse (in the CW regime, we can define an effective pulse duration and an effective repetition rate, $f_{\mathrm{rep}}$); $d = 4.4\times10^{-6}$ is the noise count probability in the single-photon detector coincidence window, which contains the dark count probability of the detector ($\approx 2\times10^{-6}$) and the spurious light count probability (luminescence noise and unfiltered ambient light); and $\eta$ is the global detection efficiency of the setup for each channel:
\begin{equation}
	\eta = \eta_{\mathrm{coll}} \, \eta_{\mathrm{det}}\,10^{-\frac{\alpha L}{10}},
\label{equationDetEff}
\end{equation}
where $\eta_{\mathrm{coll}}$ = 5\% is the collection efficiency of the setup, $\eta_{\mathrm{det}}$ = 20\% is the detection efficiency of the single-photon detectors and $\alpha$ = 0.22~dB/km is the propagation loss coefficient of each fibre spool of length $L$.\\

\noindent From the sifted and false coincidence rates, Eqs.~(\ref{equationRsift}) and (\ref{equationRfalse}), respectively, measured for zero distance and for $L=25$~km, and using $R_{\mathrm{sift}} = p_{\mathrm{true}}\,f_{\mathrm{rep}}$ and $R_{\mathrm{false}} = p_{\mathrm{false}}\,f_{\mathrm{rep}}$, we estimated the SPDC generation probability to be $\mu = 0.0035$ for an effective repetition rate $f_{\mathrm{rep}} = 78$~MHz. Finally, from the visibility $V_{\mathrm{tot}}$ measured at zero distance, we estimated a polarization measurement error $b$ = 0.06, using the relation $V_{\mathrm{tot}} = \frac{(1-2\,b)\,p_{\mathrm{true}}}{p_{\mathrm{true}}+p_{\mathrm{false}}}$ (see Eqs.~(\ref{equationQBER}) and (\ref{equationFitQBER})).\\

\noindent \textbf{Limitation of the visibility}. The polarization analysis setup at both Alice's and Bob's stations consists in a fibered half-wave plate and polarization beam-splitter with $L_f\approx 3$~m of panda-style polarization-maintaining fibers. This introduces a differential group delay $\tau_{\mathrm{PMD}} = \frac{\lambda L_f}{c L_b} \approx 3$~ps (with $L_b = 5$~mm the polarisation beat length of the panda fiber). The temporal overlap between two orthogonal polarization modes was thus limited to $\eta = \int f_A(t)f_B(\tau_{\mathrm{PMD}}-t)dt \approx$ 84\% (where $f_A(t)$ and $f_B(t)$ are the normalized temporal transmission functions of the DWDM channels). As a result, the visibility of the measured entangled state is upper bounded by $V_{\mathrm{tot}}^{\mathrm{max}} = \frac{1+\eta}{2} \approx$ 92\% \cite{Trapateau2015}.

%%%%%%%%%%%%%%%%%%%%%%%%%%%%%%%%%%%%%%%%%%%%%%%%%%%%%%%%%%%%%%%%%%%%%%
\section*{References}

\bibliography{draftQSTbib}
\bibliographystyle{iopart-num}

%\begin{thebibliography}{99}
%\end{thebibliography}

\end{document}